\documentclass{article}
\usepackage{spconf,graphicx}
\usepackage{amsmath,amssymb,amsfonts,amsthm}
\usepackage{subfig}
\usepackage{multirow}
\graphicspath{{hres/}}

\def\papertitle{MODELING NONLINEAR AUDIO EFFECTS WITH END-TO-END DEEP NEURAL NETWORKS}
\def\paperauthor{Marco A. Mart\'{i}nez Ram\'{i}rez, Joshua D. Reiss}
\title{\papertitle}
\name{\paperauthor}
\address{{Centre for Digital Music,} \\ Queen Mary University of London\\ Mile End Road, London E1 4NS, UK}
\begin{document}
\ninept
\maketitle
\begin{abstract}
In the context of music production, distortion effects are mainly used for aesthetic reasons and are usually applied to electric musical instruments. Most existing methods for nonlinear modeling are often either simplified or optimized to a very specific circuit. In this work, we investigate deep learning architectures for audio processing and we aim to find a general purpose end-to-end deep neural network to perform modeling of nonlinear audio effects. We show the network modeling various nonlinearities and we discuss the generalization capabilities among different instruments.
\end{abstract}
\section{Introduction}
\label{sec:intro}

Audio effects modeling is the process of emulating an audio effect unit and often seeks to recreate the sound of an analog reference device \cite{smith2010physical}. Correspondingly, an audio effect unit is an analog or digital signal processing system that transforms certain characteristics of the sound source. These transformations can be linear or nonlinear, with memory or memoryless. Most common audio effects' transformations are based on dynamics, such as compression; tone such as distortion; frequency such as equalization (EQ) or pitch shifters; and time such as artificial reverberation or chorus.

Nonlinear audio effects such as overdrive are widely used by musicians and sound engineers \cite{zolzer2011dafx}. These type of effects are based on the alteration of the waveform which leads to amplitude and harmonic distortion. This transformation is achieved via the nonlinear behavior of certain components of the circuitry, which apply a waveshaping nonlinearity to the audio signal amplitude in order to add harmonic and inharmonic overtones. Thus, a waveshaping transformation consists in using a nonlinear function to distort the incoming waveform into a different shape, which depends on the amplitude of the incoming signal \cite{puckette2007theory}.

Since a nonlinear element cannot be characterized by its impulse response, frequency response or transfer function \cite{smith2010physical}, digital emulation of nonlinear audio effects has been extensively researched \cite{pakarinen2009review}. 
Different methods have been proposed such as \textit{memoryless static waveshaping} \cite{moller2002measurement, santagata2007non}, where system-identification methods are used in order to model the nonlinearity; \textit{dynamic nonlinear filters} \cite{karjalainen2006virtual}, where the waveshaping curve changes its shape as a function of system-state variables; \textit{analytical methods} \cite{abel2006technique, helie2006use}, where the nonlinearity is linearized via Volterra series theory or black-box modeling such as Wiener and Hammerstein models \cite{eichas2018virtual, gilabert2005wiener}; and \textit{circuit simulation techniques} \cite{yeh2008numerical, yeh2008simulating, yeh2010automated}, where nonlinear filters are derived from the differential equations that describe the circuit. Recurrent neural networks have been explored as preliminary studies in \cite{covert2013vacuum,schmitz2018nonlinear, zhang2018vacuum}, where the proposed models may require a more extensive evaluation.

In order to achieve optimal results, these methods are often either greatly simplified or highly optimized to a very specific circuit. Thus, without resorting to further complex analysis methods or prior knowledge about the circuit, it is difficult to generalize the methods among different audio effects. This lack of generalization is accentuated when we consider that each unit of audio effects is also composed of components other than the nonlinearity. These components also need to be modeled and often involve filtering before and after the waveshaping, as well as hysteresis or attack and release gates.

\begin{figure*}[ht]
\center
\makebox[0pt]{\includegraphics[width=1.075\linewidth]{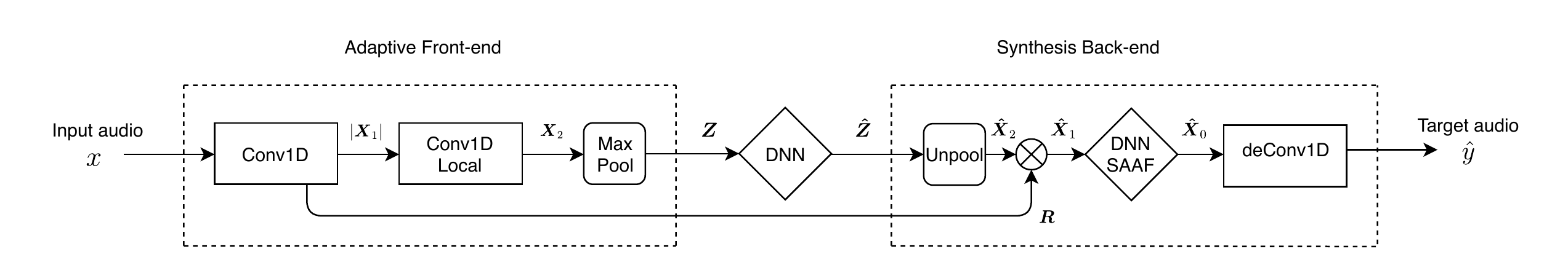}}
\caption{\label{fig:model}{Block diagram of the proposed model; adaptive front-end, synthesis back-end and latent-space DNN. }}
\end{figure*}

End-to-end learning corresponds to the integration of an entire problem as a single indivisible task that must be learned from \textit{end-to-end}. The desired output is obtained from the input by learning directly from the data \cite{muller2006off}. Deep learning architectures using this principle have experienced significant growth in music information retrieval \cite{pons2017end, venkataramani2017adaptive}, since by learning directly from raw audio, the amount of required prior knowledge is reduced and the engineering effort is minimized \cite{dieleman2014end}. 

End-to-end deep neural networks (DNN) for audio processing have been implemented in \cite{martinez2018end}, where EQ modeling was achieved with convolutional neural networks (CNN). We build on this model in order to emulate much more complex transformations such as nonlinearities. To the best of our knowledge, prior to this work, deep learning architectures has not been successfully implemented to model nonlinear and linear audio effects. 

We explore nonlinear emulation as a content-based transformation without explicitly obtaining the solution of the nonlinear system. We show the model performing nonlinear modeling for \textit{distortion}, \textit{overdrive}, \textit{amplifier emulation} and \textit{combinations of linear and nonlinear} audio effects. 

\section{Methods}
\label{sec:methods}

\vspace{-3ex}\subsection{Model}

The model is entirely based on the time-domain and is divided into three parts: adaptive front-end, synthesis back-end and latent-space DNN. We build on the model from \cite{martinez2018end} and we incorporate a new layer into the synthesis back-end. The model is depicted in Fig. \ref{fig:model}, and may seem similar to the nonlinear system measurement technique from \cite{abel2006technique}, as it is based on a parallel combination of the cascade of input filters, memoryless nonlinearities, and output filters.

The \textbf{adaptive front-end} consist of a convolutional encoder. It contains two CNN layers, one pooling layer and one residual connection. The front-end performs time-domain convolutions with the raw audio in order to map it into a latent-space. It also generates a residual connection which facilitates the reconstruction of the waveform by the back-end.

The input layer has $128$ one-dimensional filters of size $64$ and is followed by the \textit{absolute value} as nonlinear activation function. The second layer has $128$ filters of size $128$ and each filter is locally connected. This means we follow a filter bank architecture since each filter is only applied to its corresponding row in $|\boldsymbol{X}_{1}|$ and we also decrease significantly the number of trainable parameters. This layer is followed by the \textit{softplus} nonlinearity. 

From Fig. \ref{fig:model}, $\boldsymbol{R}$ is the matrix of the residual connection, $\boldsymbol{X}_{1}$ is the feature map or frequency decomposition matrix after the input signal $x$ is convolved with the kernel matrix $\boldsymbol{W}_{1}$, and $\boldsymbol{X}_{2}$ is the second feature map obtained after the local convolution with $\boldsymbol{W}_{2}$, the kernel matrix of the second layer. The \textit{max-pooling} layer is a moving window of size $16$, where positions of maximum values are stored and used by the back-end. Also, in the front-end, we include a batch normalization layer before the max-pooling operation.

The \textbf{latent-space DNN} contains two layers. Following the filter bank architecture, the first layer is based on locally connected dense layers of $64$ hidden units and the second layer consists of a fully connected layer of $64$ hidden units. Both of these layers are followed by the \textit{softplus} function. Since $\boldsymbol{Z}$ corresponds to a latent representation of the input audio. The DNN modifies this matrix into a new latent representation $\boldsymbol{\hat{Z}}$ which is fed into the synthesis back-end. Thus, the front-end and latent-space DNN carry out the input filtering operations of the given nonlinear task.

The \textbf{synthesis back-end} inverts the operations carried out by the front-end and applies various dynamic nonlinearities to the modified frequency decomposition of the input audio signal $\hat{X}_{1}$. Accordingly, the back-end consists of an unpooling layer, a deep neural network with smooth adaptive activation functions (DNN-SAAF) and a single CNN layer.

\underline{DNN-SAAF:} These consist of four fully connected dense layers of $128$, $64$, $64$ and $128$ hidden units respectively. All dense layers are followed by the \textit{softplus} function with the exception of the last layer. Since we want the network to learn various nonlinear filters for each row of $\hat{X}_{1}$, we use locally connected Smooth Adaptive Activation Functions (SAAF) \cite{hou2017convnets} as the nonlinearity for the last layer. 

SAAFs consist of piecewise second order polynomials which can approximate any continuous function and are regularized under a Lipschitz constant to ensure smoothness. It has been shown that the performance of CNNs in regression tasks has improved when adaptive activation functions have been used \cite{hou2017convnets}, as well as their generalization capabilities and learning process timings \cite{solazzi2000artificial, uncini2003audio, godfrey2015continuum}.

We tested different types of adaptive activation functions, such as parametric hyperbolic tangent, parametric sigmoid and fifth order polynomials. Nevertheless, we found stability problems and non optimal results when modeling complex nonlinearities.

The back-end accomplishes the reconstruction of the target audio signal by the following steps. First, a discrete approximation $\boldsymbol{\hat{X}}_{2}$ is obtained by upsampling $\boldsymbol{Z}$ at the locations of the maximum values from the pooling operation. Then the approximation $\boldsymbol{\hat{X}}_{1}$ of matrix $\boldsymbol{X}_{1}$ is obtained through the element-wise multiplication of the residual $\boldsymbol{R}$ and $\boldsymbol{\hat{X}}_{2}$. In order to obtain $\hat{X}_{0}$, the nonlinear filters from DNN-SAAF are applied to $\hat{X}_{1}$. Finally, the last layer corresponds to the deconvolution operation, which can be implemented by transposing the first layer transform. 

We train two types of models: \textit{model-1} without dropout layers within the dense layers of the latent-space DNN and DNN-SAAF, and \textit{model-2} with dropout layers among the hidden units of these layers. All convolutions are along the time dimension and all strides are of unit value. This means, during convolution, we move the filters one sample at a time. The models have approximately $600k$ trainable parameters, which represents a model that is not very large or difficult to train.

\begin{figure*}[ht]
\begin{minipage}[b]{.25\textwidth}
  \centering
  \centerline{\includegraphics[width=1.\linewidth,height=0.12\textheight]{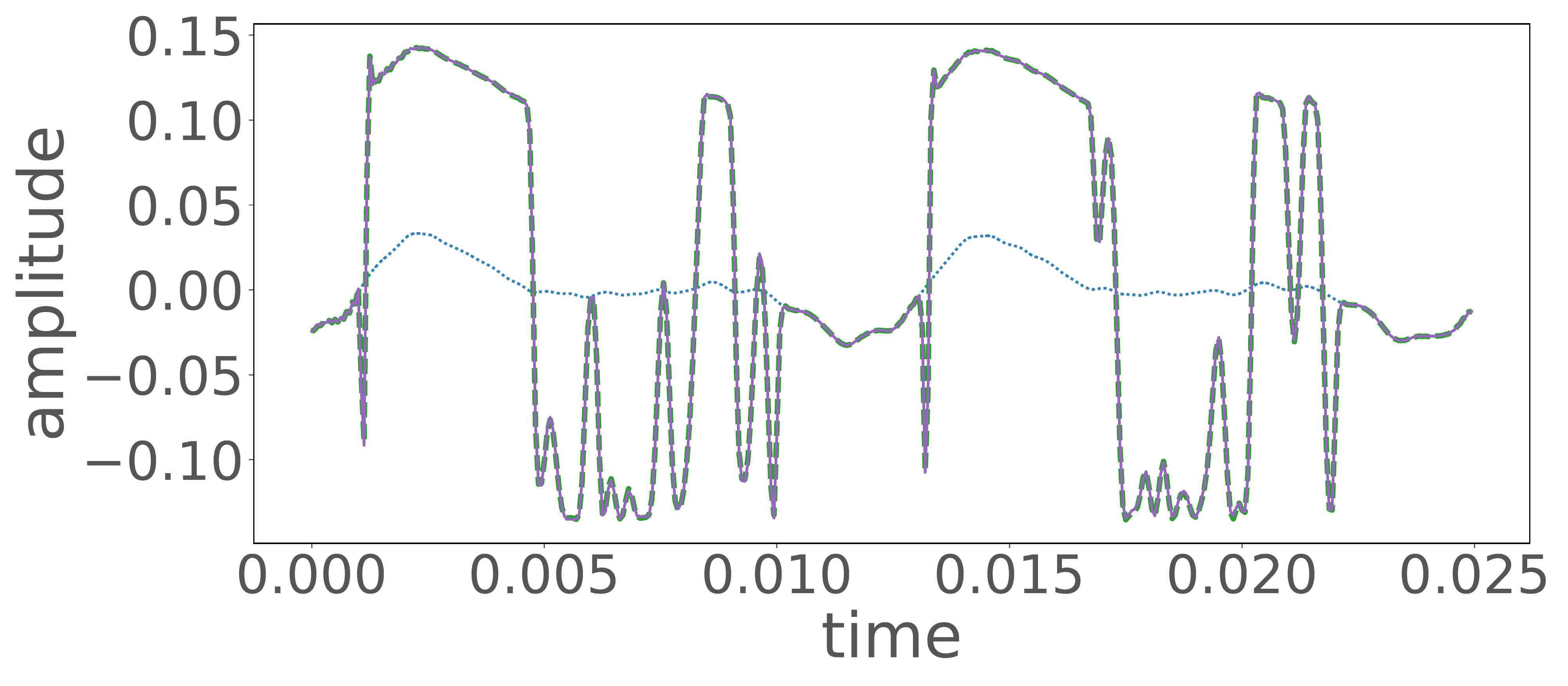}\label{fig:distortion-1-frames}}
  \centerline{\includegraphics[width=1.\linewidth,height=0.12\textheight]{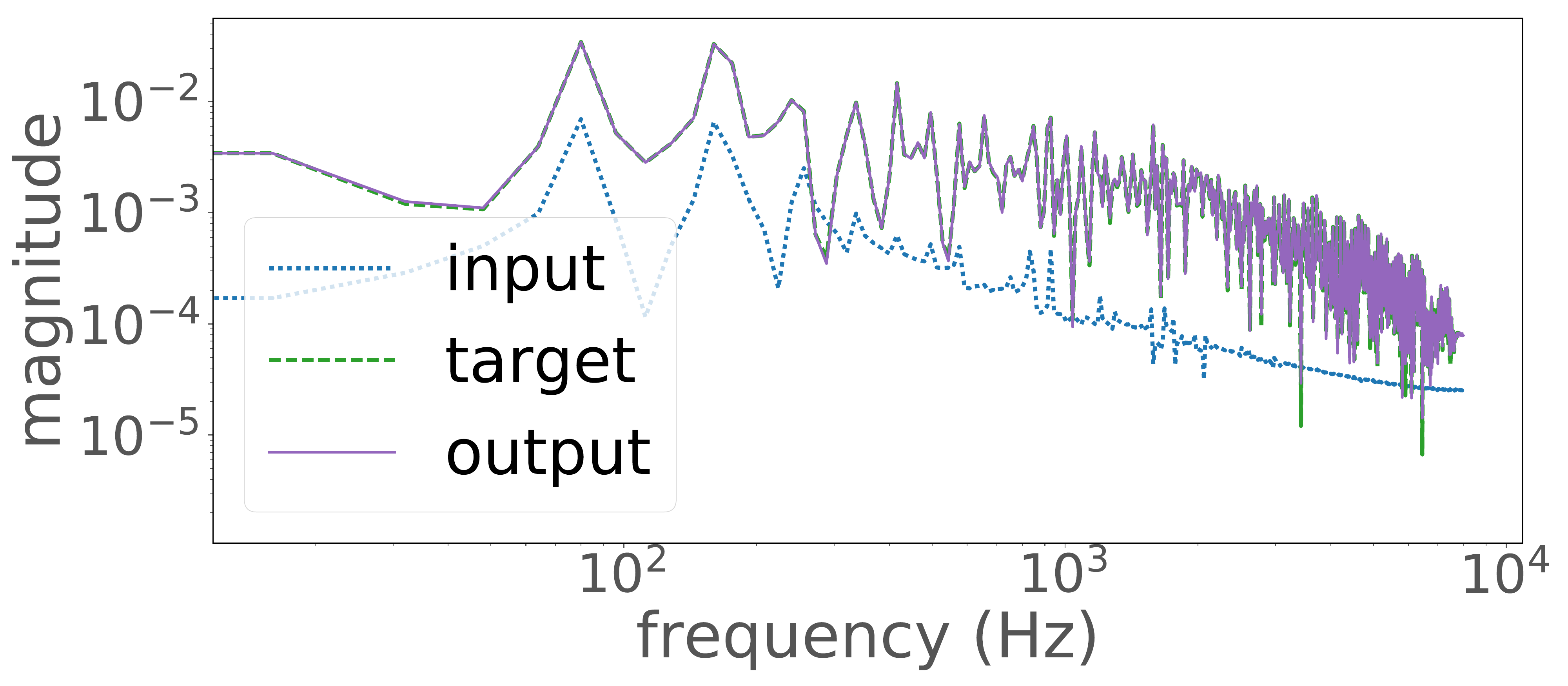}\label{fig:distortion-1-fft}}
%  \vspace{1.5cm}
  \centerline{(a)}
\end{minipage}
% \hfill
\begin{minipage}[b]{0.24\textwidth}
  \centering
  \centerline{\includegraphics[width=1.\linewidth,height=0.2475\textheight]{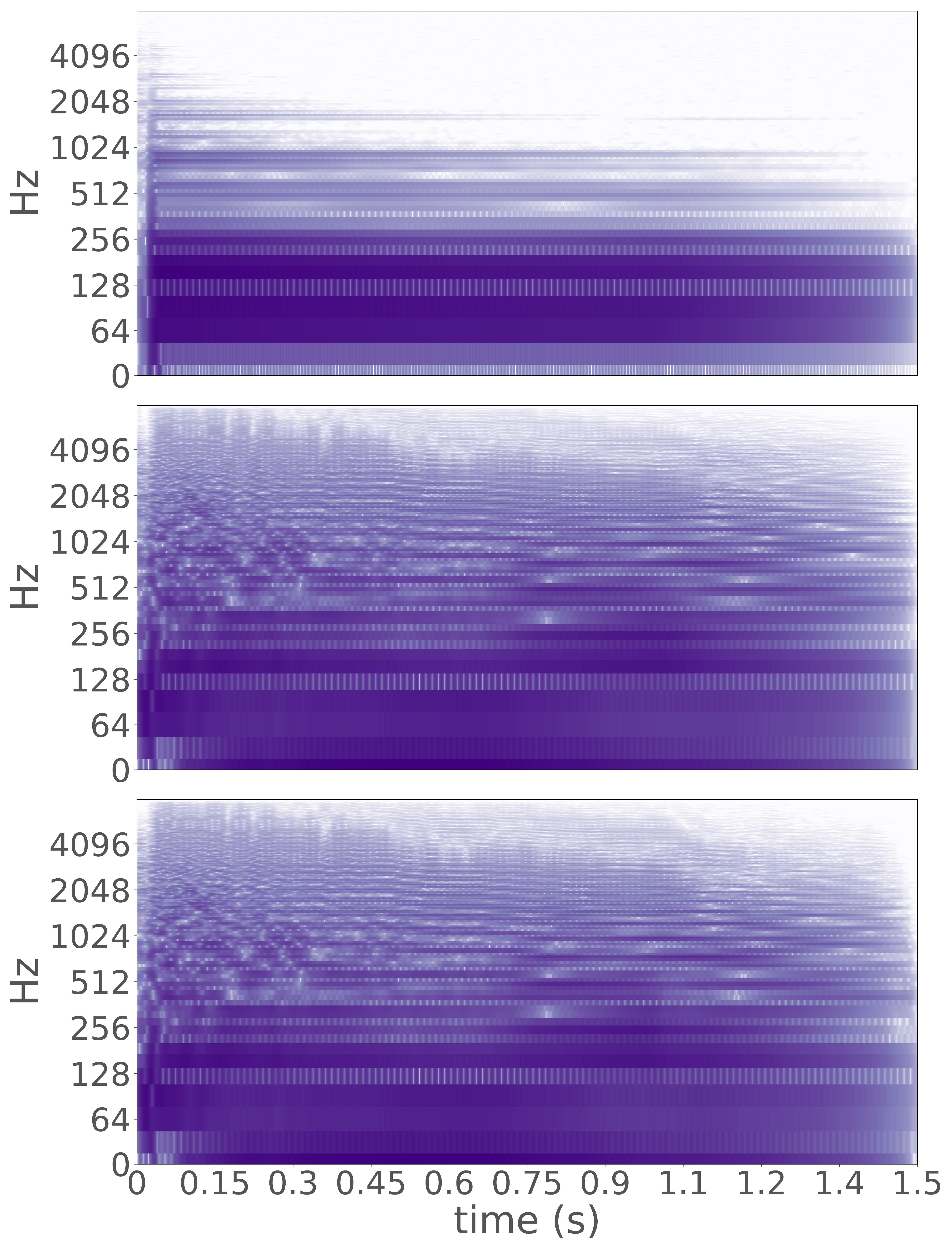}\label{fig:distortion-1-spectogram}}
%  \vspace{1.5cm}
  \centerline{(b) }
\end{minipage}
\begin{minipage}[b]{.25\textwidth}
  \centering
  \centerline{\includegraphics[width=1.\linewidth,height=0.12\textheight]{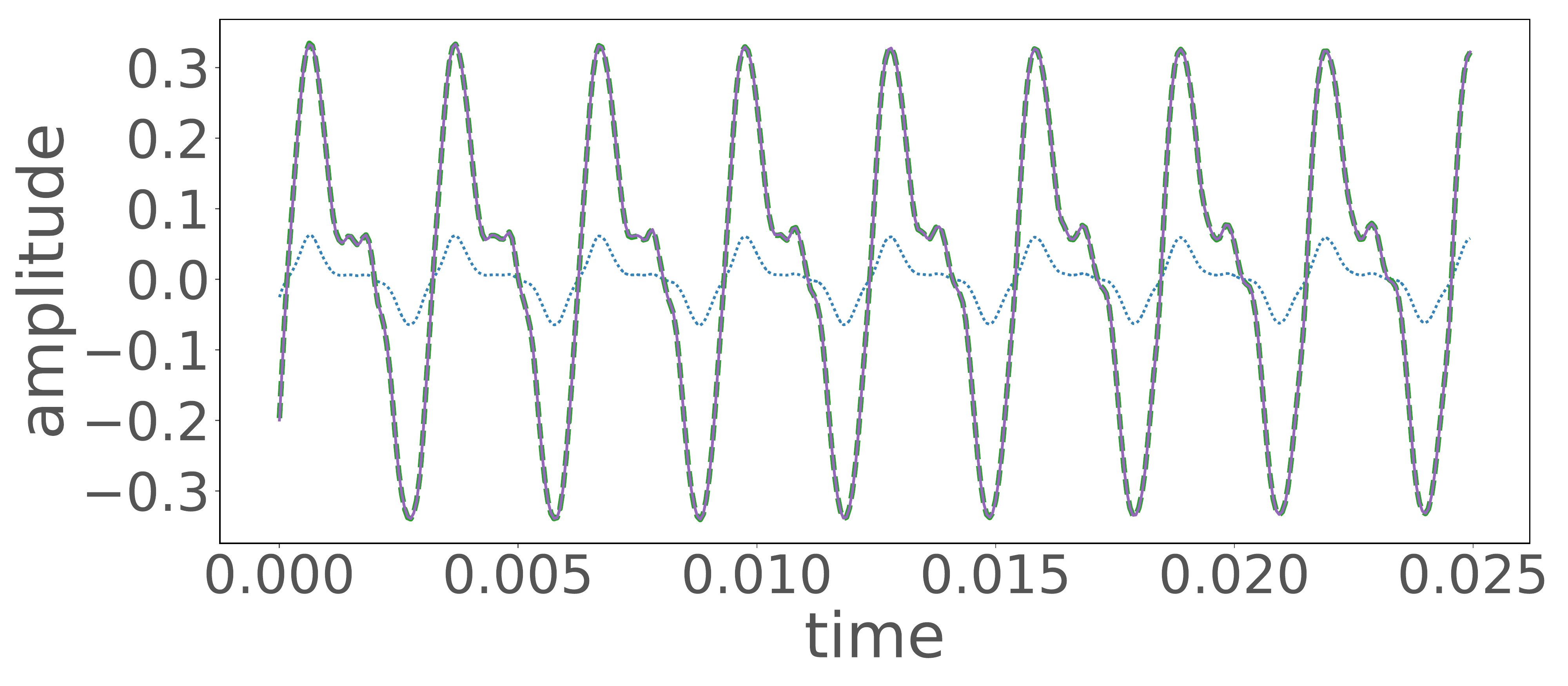}\label{fig:distortion-2-frames}}
  \centerline{\includegraphics[width=1.\linewidth,height=0.12\textheight]{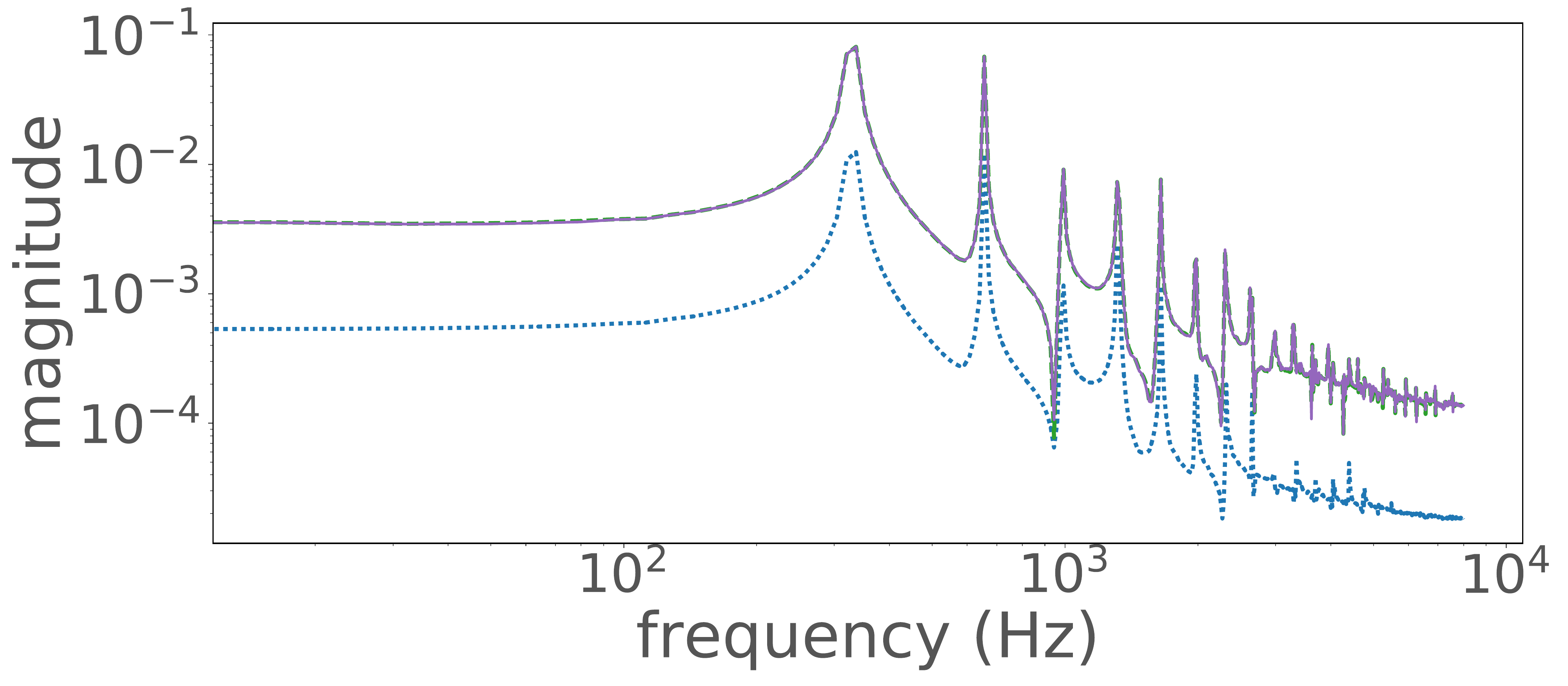}\label{fig:distortion-2-fft}}
%  \vspace{1.5cm}
  \centerline{(c)}
\end{minipage}
% \hfill
\begin{minipage}[b]{0.24\textwidth}
  \centering
  \centerline{\includegraphics[width=1.\linewidth,height=0.2475\textheight]{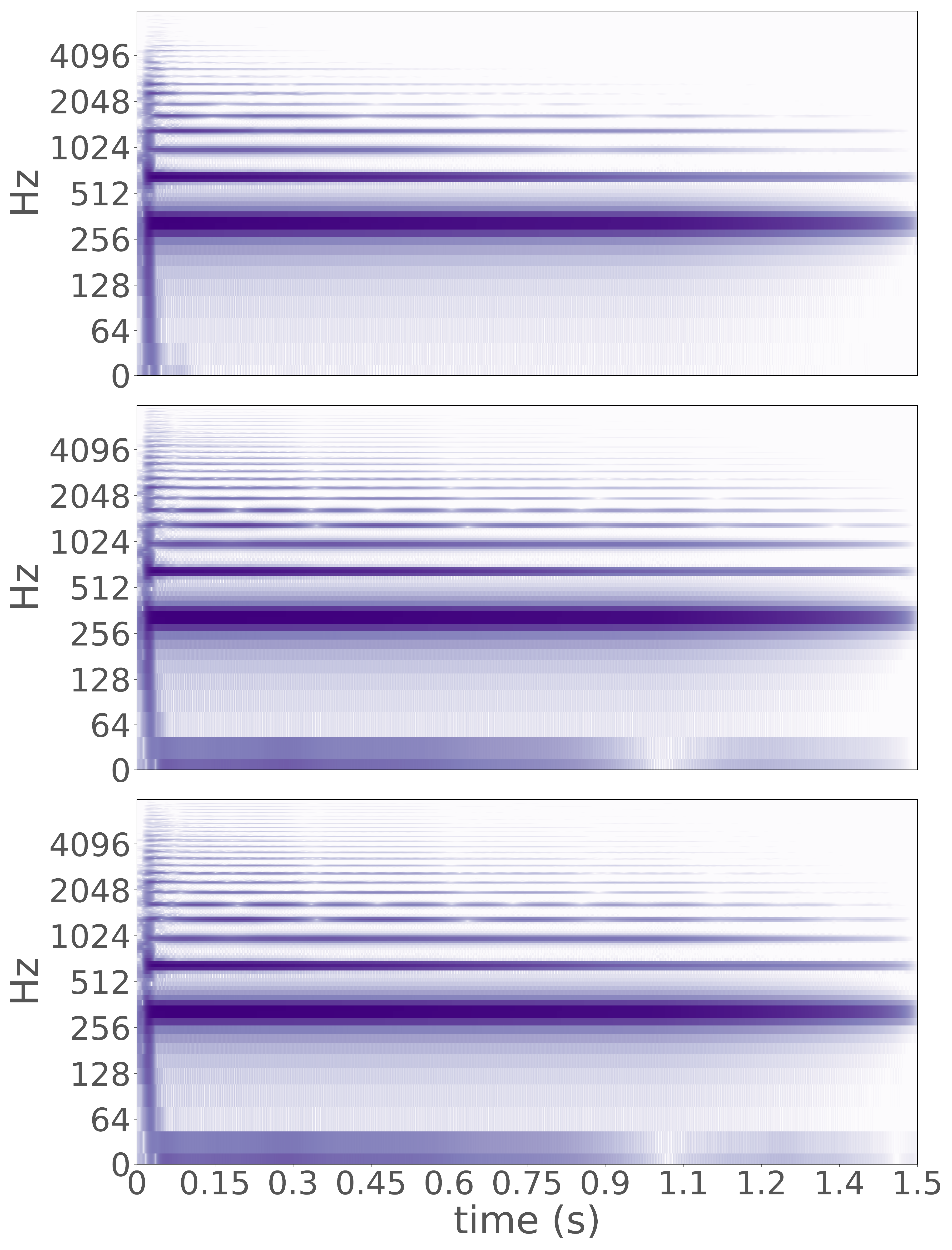}\label{fig:distortion-2-spectogram}}
%  \vspace{1.5cm}
  \centerline{(d)}
\end{minipage}
\caption{Results with the test dataset for \ref{fig:frames}a-b) \textit{model-1} bass guitar distortion setting \# 1, and \ref{fig:frames}c-d) \textit{model-2} electric guitar overdrive setting \# 2. A segment of the input, target and output frames and their respective FFT magnitudes is shown. Also, from top to bottom: input, target and output spectrograms of the test samples; color intensity represents higher magnitude.}
\label{fig:frames}
\end{figure*}

\begin{figure*}[ht]
\begin{minipage}[b]{.245\textwidth}
  \centering
  \centerline{\includegraphics[width=1.\linewidth,height=0.12\textheight]{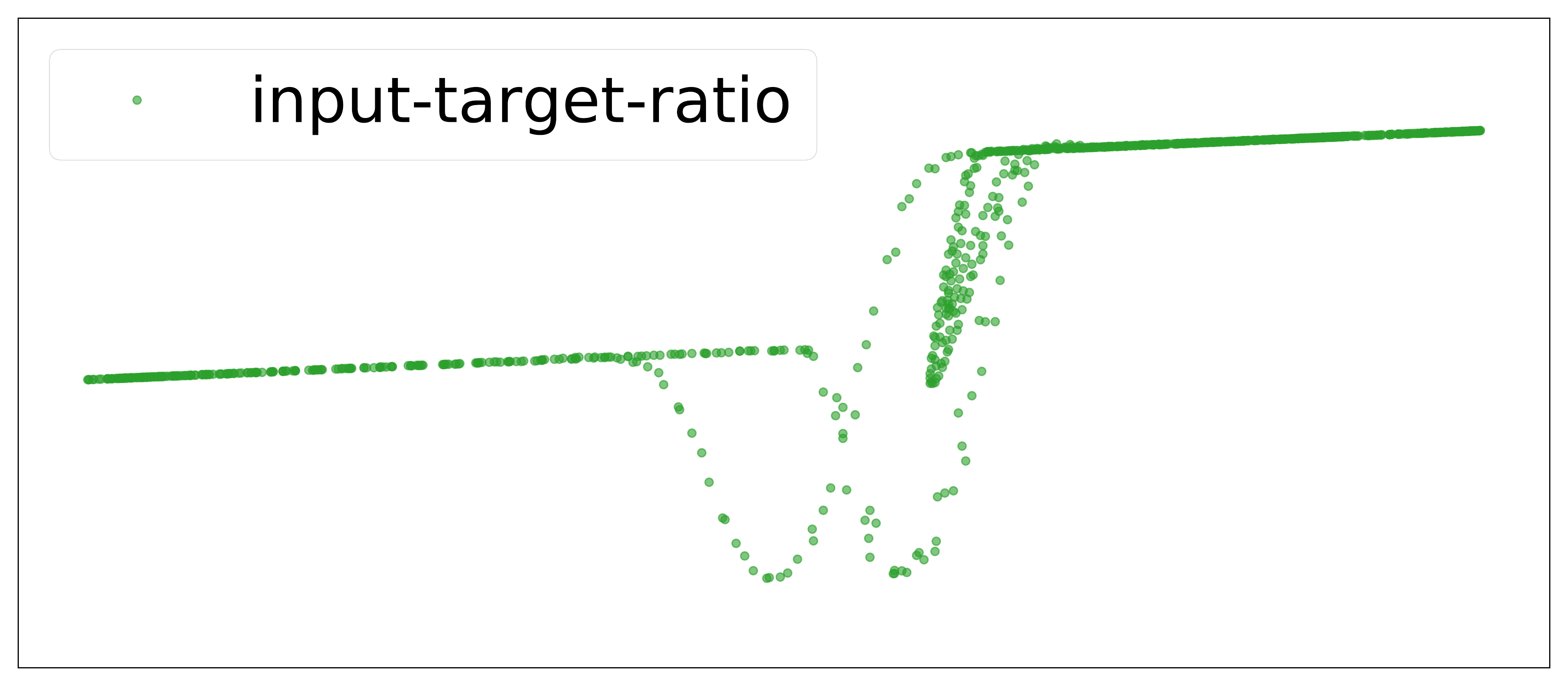}\label{fig:distortion-1-ratio1}}
  \centerline{\includegraphics[width=1.\linewidth,height=0.12\textheight]{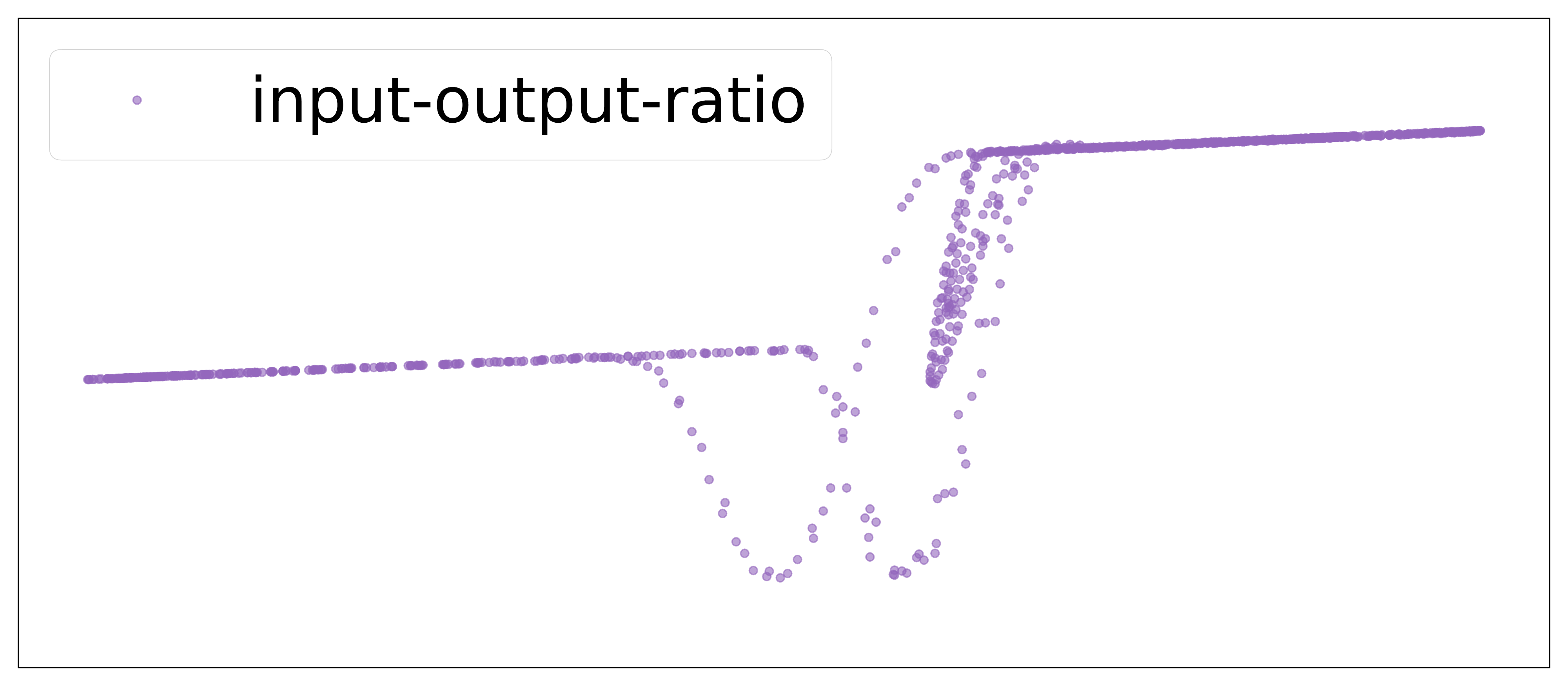}\label{fig:distortion-1-ratio2}}
%  \vspace{1.5cm}
  \centerline{(a)}
\end{minipage}
% \hfill
\begin{minipage}[b]{0.245\textwidth}
  \centering
    \centerline{\includegraphics[width=1.\linewidth,height=0.12\textheight]{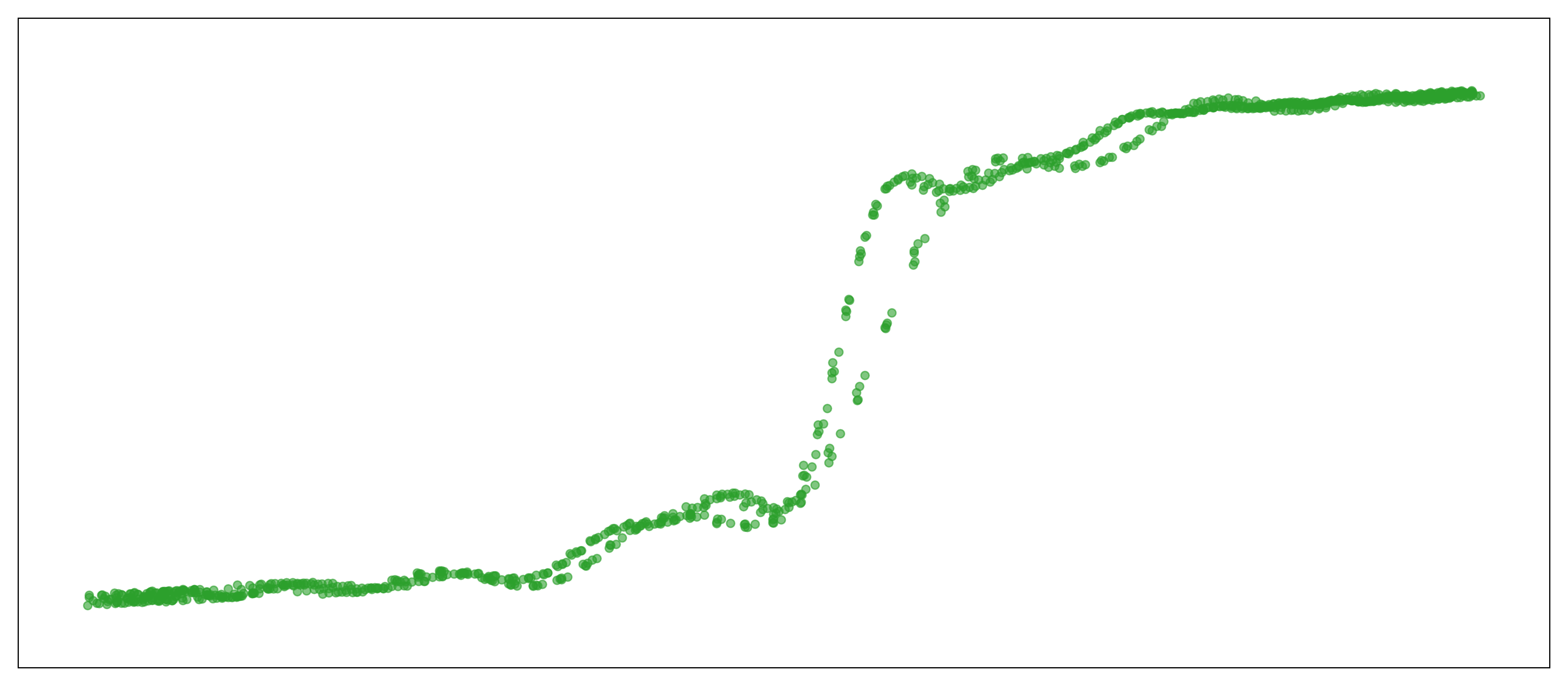}\label{fig:distortion-2-ratio1}}
  	\centerline{\includegraphics[width=1.\linewidth,height=0.12\textheight]{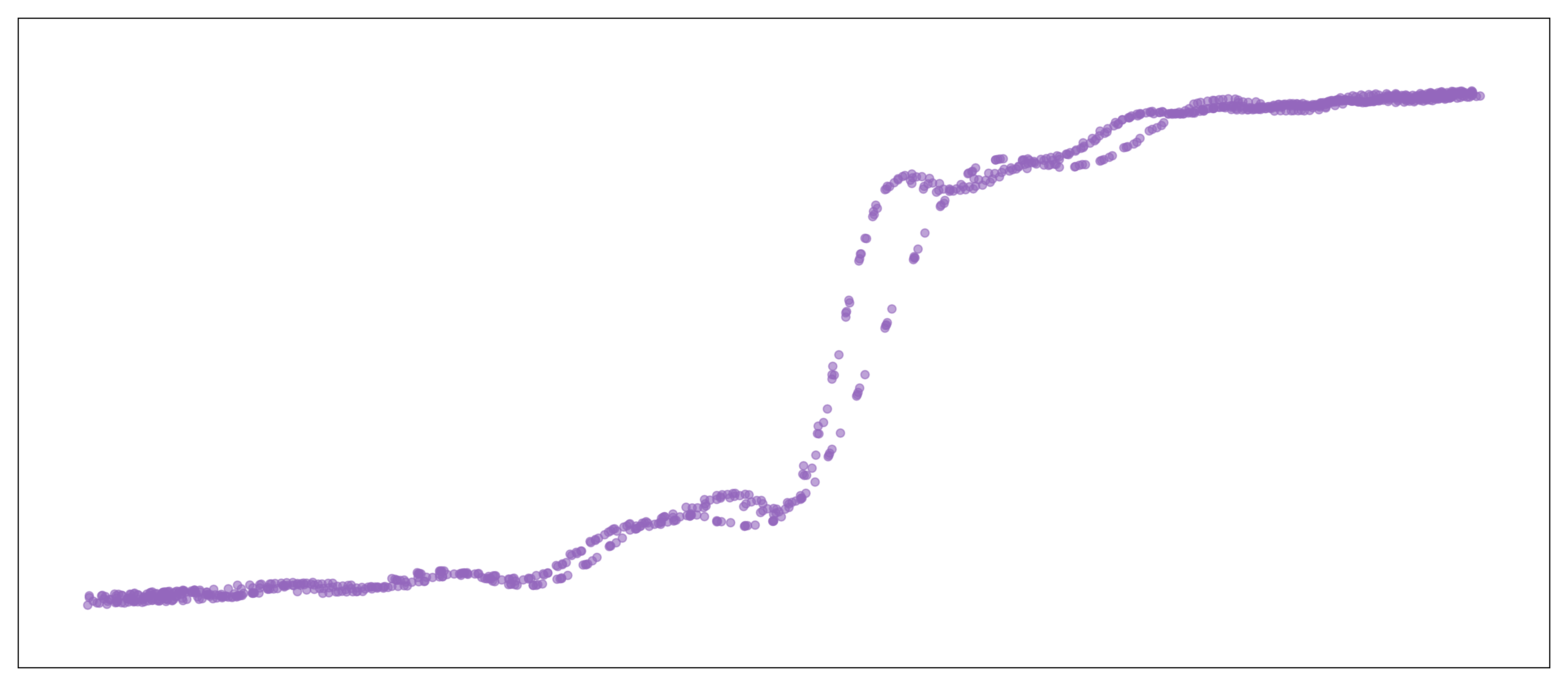}\label{fig:distortion-2-ratio2}}
%  \vspace{1.5cm}
  \centerline{(b)}
\end{minipage}
\begin{minipage}[b]{.245\textwidth}
  \centering
  \centerline{\includegraphics[width=1.\linewidth,height=0.12\textheight]{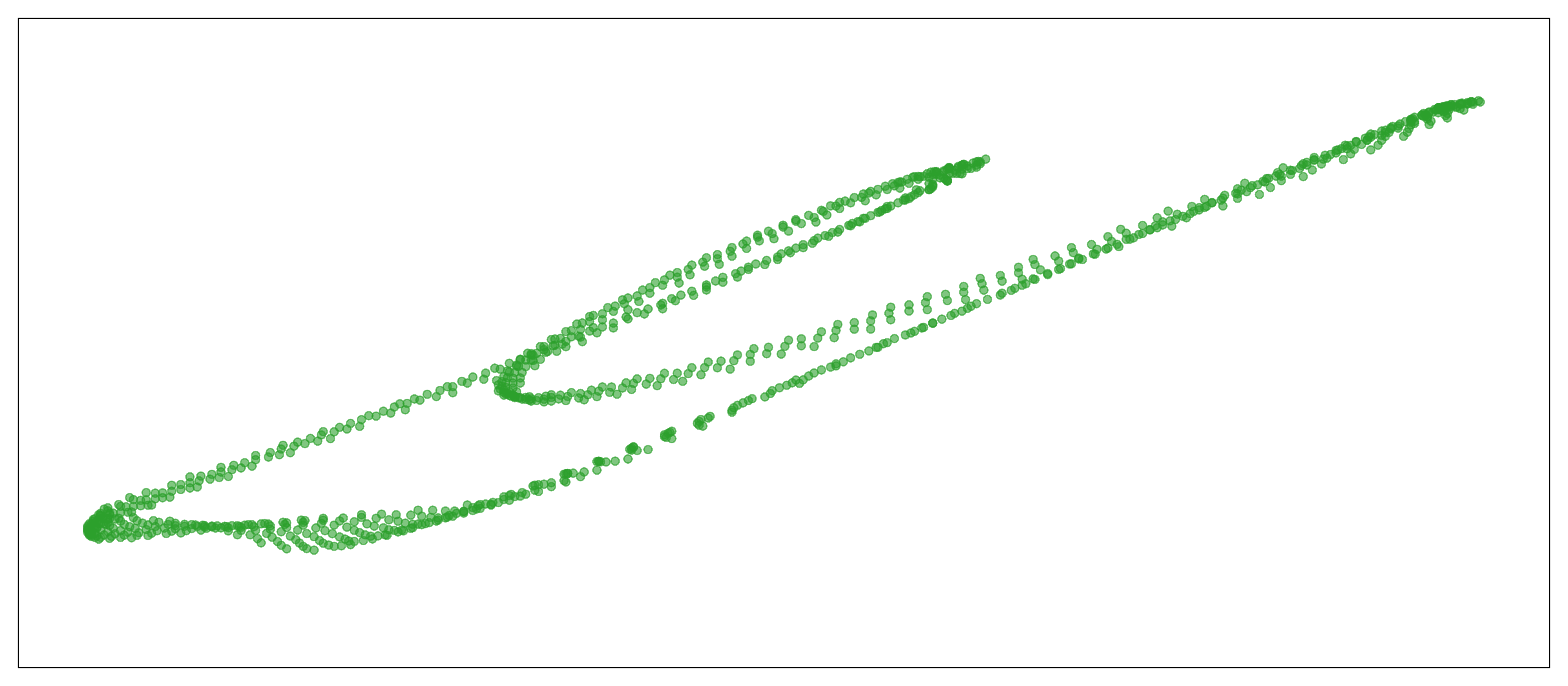}\label{fig:overdrive-1-ratio1}}
  \centerline{\includegraphics[width=1.\linewidth,height=0.12\textheight]{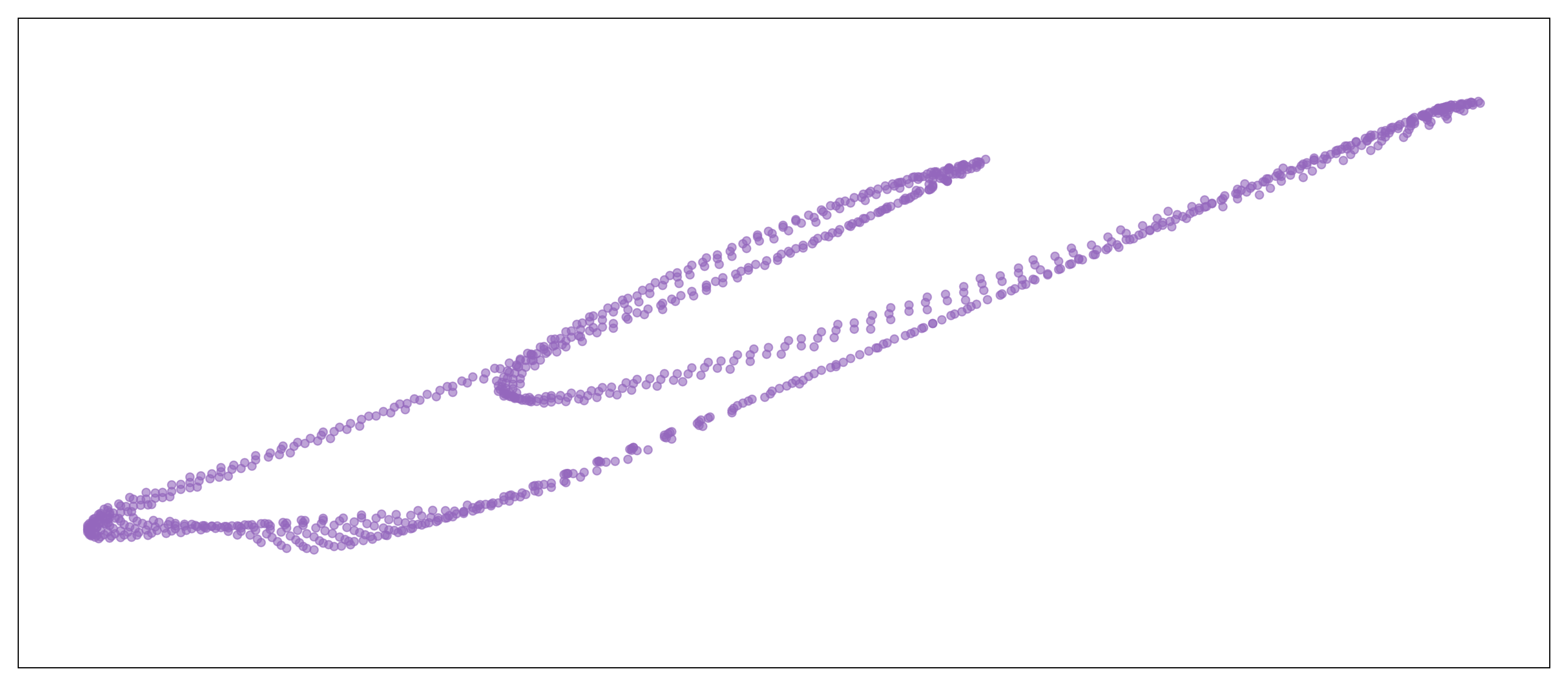}\label{fig:overdrive-1-ratio2}}
%  \vspace{1.5cm}
  \centerline{(c)}
\end{minipage}
% \hfill
\begin{minipage}[b]{0.245\textwidth}
  \centering
  \centerline{\includegraphics[width=1.\linewidth,height=0.12\textheight]{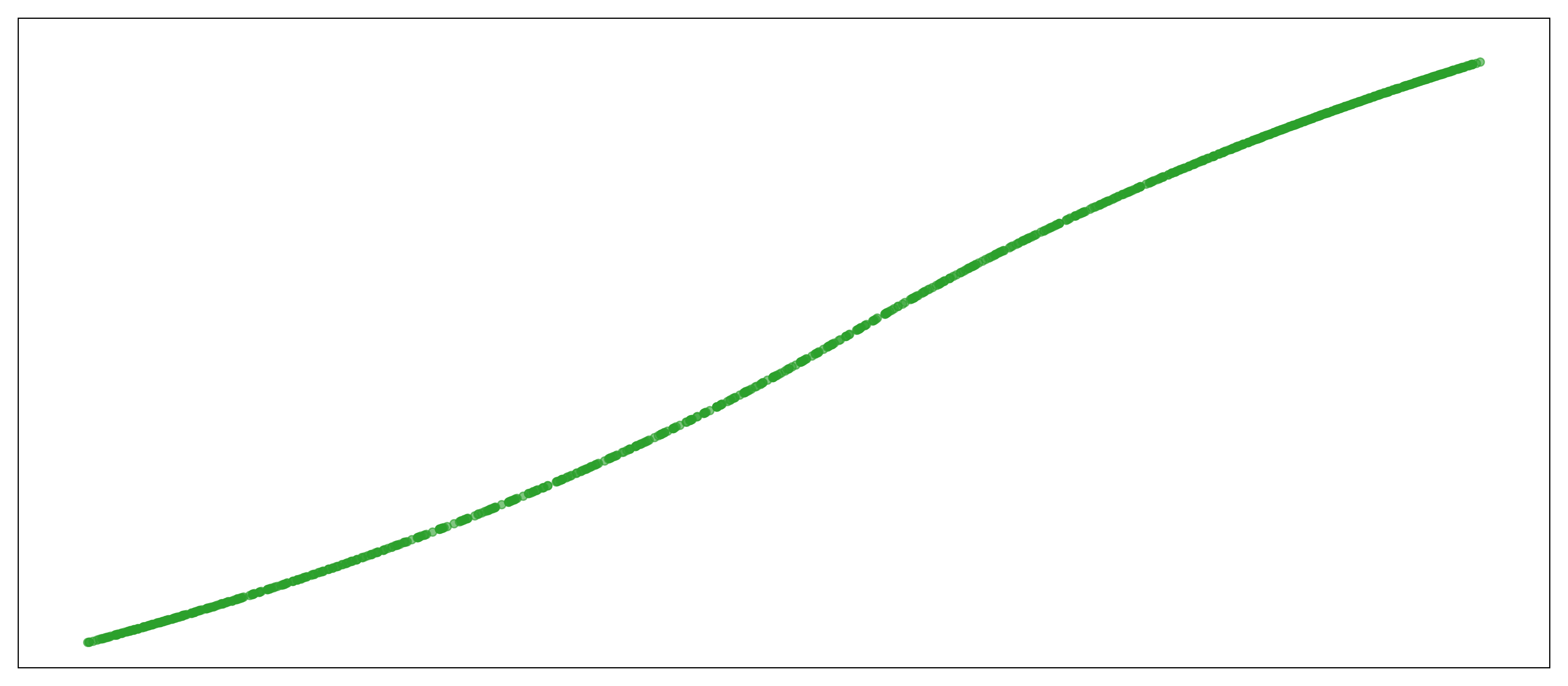}\label{fig:overdrive-2-ratio1}}
  \centerline{\includegraphics[width=1.\linewidth,height=0.12\textheight]{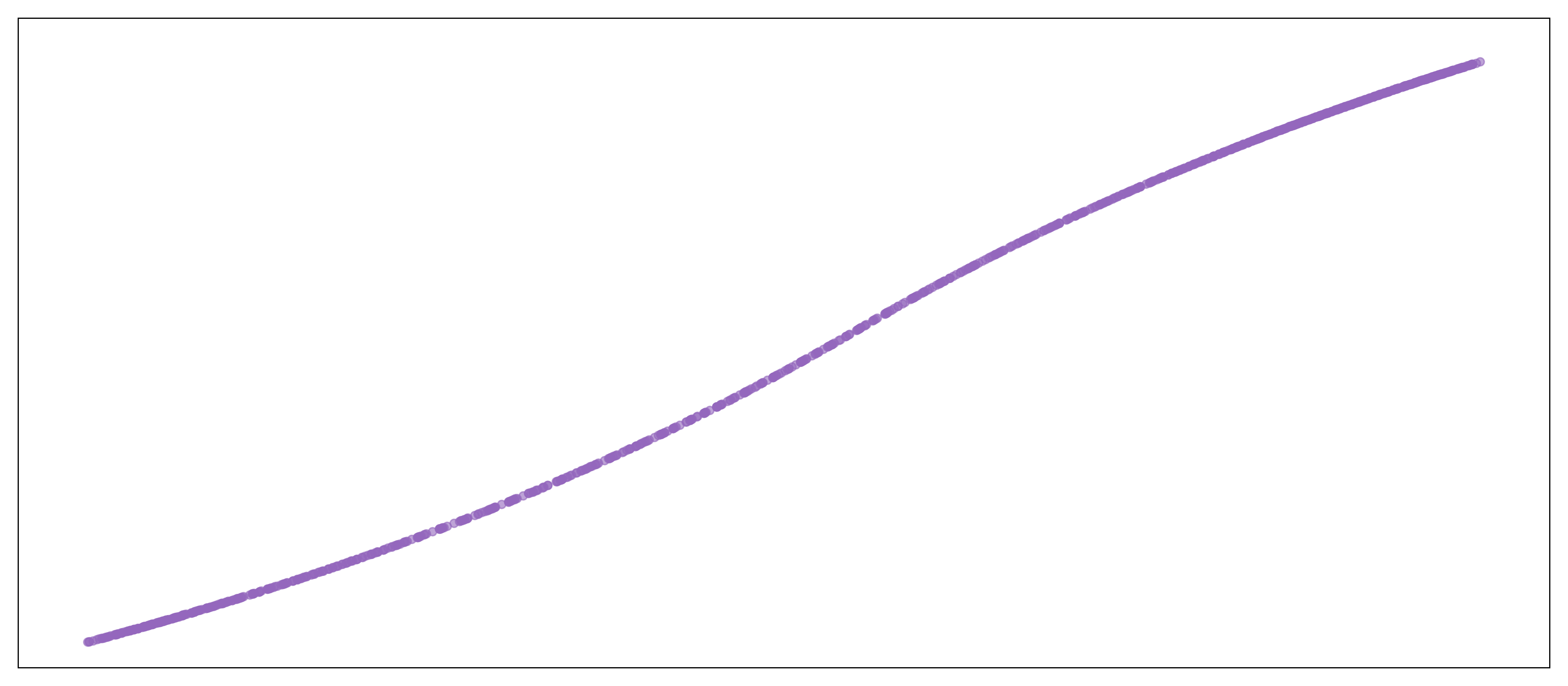}\label{fig:overdrive-2-ratio2}}
%  \vspace{1.5cm}
  \centerline{(d)}
\end{minipage}
\caption{Input-Target and Input-Output waveshaping curve for selected settings. \ref{fig:ratios}a) \textit{model-1} bass guitar distortion task \#1. \ref{fig:ratios}b) \textit{model-1} electric guitar distortion setting \#2. \ref{fig:ratios}c) \textit{model-2} bass guitar overdrive setting \#1. \ref{fig:ratios}d) \textit{model-2} electric guitar overdrive setting \#2. X-axis is input amplitude and Y-axis is target/output amplitude.}
\label{fig:ratios}
\end{figure*}
Based on end-to-end deep neural networks, we introduce a general purpose deep learning architecture for modeling nonlinear audio effects. Thus, for an arbitrary combination of linear and nonlinear memoryless audio effects, the model learns how to process the audio directly in order to match the target audio. Given a nonlinearity, consider $x$ and $y$ the raw and distorted audio signals respectively. In order to obtain a $\hat{y}$ that matches the target $y$, we train a deep neural network to modify $x$ based on the nonlinear task.

% \vspace{-3ex}
\subsection{Training}

The training of the model is performed in two steps. The first step is to train only the convolutional layers for an unsupervised learning task, while the second step is within a supervised learning framework for the entire network. During the first step only the weights of \textit{Conv1D} and \textit{Conv1D-Local} are optimized and both the raw audio $x$ and distorted audio $y$ are used as input and target functions. 

Once the model is pretrained, the latent-space DNN and DNN-SAAF are incorporated into the model, and all the weights of the convolutional and dense layers are updated. The loss function to be minimized is the mean absolute error ($mae$) between the target and output waveforms. In both training procedures the input and target audio are sliced into frames of $1024$ samples with hop size of $64$ samples. The mini-batch was $32$ frames and $1000$ iterations were carried out for each training step. 

% \vspace{-3ex}
\subsection{Dataset}

The audio is obtained from the \textit{IDMT-SMT-Audio-Effects} dataset \cite{stein2010automatic}, which corresponds to individual 2-second notes and covers the common pitch range of various 6-string electric guitars and 4-string bass guitars. 

The recordings include the raw notes and their respective effected versions after $3$ different settings for each effect. We use unprocessed and processed audio with distortion, overdrive, and EQ. In addition, we also apply a custom audio effects chain (FxChain) to the raw audio. The FxChain consist of a lowshelf filter ($gain=+20$dB) followed by a highshelf filter ($gain=-20$dB) and an overdrive ($gain=+30$dB). Both filters have a cut-off frequency of $500$ Hz. Three different configurations were explored by placing the overdrive as the last, second and first effect of the cascade. 

We use $624$ raw and distorted notes for each audio effect setting. The test and validation notes correspond to 10\% of this subset and contain recordings of a different electric guitar and bass guitar. In order to reduce training times, the recordings were downsampled to $16$ kHz, however, the model could be trained with higher sampling rates.
% \vspace{-3ex}
\section{Results \& Analysis}
\label{sec:results}

\begin{table}[t]
\begin{center}
\caption{\textit{mae} values of the bass guitar and electric guitar models with the test datasets. }

% \resizebox{1.\columnwidth}{!}{
% \renewcommand{\arraystretch}{1.0}
\begin{tabular}[c] { |p{1.3cm}|p{0.2cm}|
                    |p{1.15cm}|p{1.15cm}|
                    p{1.15cm}|p{1.15cm}|}
\hline
 %& & \multicolumn{3}{c|}{ } \\%[2ex]

 \multirow{2}{1em}{Fx} & \multirow{2}{0.1pt}{$\#$} 
 &\multicolumn{2}{c|}{\vspace{0pt}Bass} &\multicolumn{2}{c|}{\vspace{0pt}Guitar}\\[1ex]
 \cline{3-6}
 & &\textit{model-1} & \textit{model-2} & \textit{model-1} & \textit{model-2} \\[1ex]
 \hline
\vspace{2pt}

\multirow{3}{1em}{Distortion}  
&\vspace{0pt} 1 
&\vspace{0pt} 0.00318 &\vspace{0pt} 0.00530
&\vspace{0pt} 0.00459 &\vspace{0pt} 0.00331\\

&\vspace{0pt} 2 
&\vspace{0pt} 0.00263 &\vspace{0pt} 0.00482
&\vspace{0pt} 0.00366 &\vspace{0pt} 0.00428\\

&\vspace{0pt} 3 
&\vspace{0pt} 0.00123 &\vspace{0pt} 0.00396
&\vspace{0pt} 0.00121 &\vspace{0pt} 0.00586 \\
\hline

\multirow{3}{1em}{Overdrive}  
&\vspace{0pt} 1 
&\vspace{0pt} 0.00040 &\vspace{0pt} 0.00437
&\vspace{0pt} 0.00066 &\vspace{0pt} 0.00720\\

&\vspace{0pt} 2 
&\vspace{0pt} 0.00011 &\vspace{0pt} 0.00131
&\vspace{0pt} 0.00048 &\vspace{0pt} 0.00389\\

&\vspace{0pt} 3 
&\vspace{0pt} 0.00037 &\vspace{0pt} 0.00206
&\vspace{0pt} 0.00072 &\vspace{0pt} 0.00436 \\
\hline

\multirow{2}{1em}{EQ}  
&\vspace{0pt} 1 
&\vspace{0pt} 0.00493 &\vspace{0pt} 0.00412
&\vspace{0pt} 0.00842 &\vspace{0pt} 0.00713\\

&\vspace{0pt} 2 
&\vspace{0pt} 0.00543 &\vspace{0pt} 0.00380 %0.00596 -> SAAF3
&\vspace{0pt} 0.00522 &\vspace{0pt} 0.00543\\
\hline

\multirow{3}{1em}{FxChain}  
&\vspace{0pt} 1 
&\vspace{0pt} 0.01171 &\vspace{0pt} 0.02103
&\vspace{0pt} 0.01421 &\vspace{0pt} 0.01423\\

&\vspace{0pt} 2 
&\vspace{0pt} 0.01307 &\vspace{0pt} 0.01365
&\vspace{0pt} 0.01095 &\vspace{0pt} 0.00957\\

&\vspace{0pt} 3 
&\vspace{0pt} 0.01380 &\vspace{0pt} 0.01773
&\vspace{0pt} 0.01778 &\vspace{0pt} 0.01396 \\
\hline

\end{tabular}
% }
\label{table:bass-losses}
\end{center}
\end{table}

\begin{table}[h!]
\begin{center}
\caption{Evaluation of the generalization capabilities of the models. \textit{mae} values for \textit{model-1} and \textit{model-2} when tested with a different instrument recording and with the NSynth test dataset.}
% \resizebox{1.\columnwidth}{!}{
% \renewcommand{\arraystretch}{1.0}
\begin{tabular}[c] { |p{1.3cm}|p{0.2cm}|
                    |p{1.15cm}|p{1.15cm}|
                    p{1.15cm}|p{1.15cm}|}
\hline

 \multirow{2}{1em}{Fx} & \multirow{2}{0.1pt}{$\#$} 
 &\multicolumn{2}{c|}{\vspace{0pt}Bass} &\multicolumn{2}{c|}{\vspace{0pt}Guitar}\\[1ex]
 \cline{3-6}
 & &\textit{model-1} & \textit{model-2} & \textit{model-1} & \textit{model-2} \\[1ex]
 \hline
\vspace{2pt}

\multirow{3}{1em}{\vspace{0.1ex} FxChain-different instrument}  
&\vspace{0pt} 1 
&\vspace{0pt} 0.02235 &\vspace{0pt} 0.01670 
&\vspace{0pt} 0.10375 &\vspace{0pt} 0.09501\\

&\vspace{0pt} 2 
&\vspace{0pt} 0.02153 &\vspace{0pt} 0.01374
&\vspace{0pt} 0.06705 &\vspace{0pt} 0.06397\\

&\vspace{0pt} 3 
&\vspace{0pt} 0.02936 &\vspace{0pt} 0.02072
&\vspace{0pt} 0.10900 &\vspace{0pt} 0.10254 \\
\hline
\multirow{3}{1em}{\vspace{1ex} FxChain-NSynth}  
&\vspace{0pt} 1 
&\vspace{0pt} 0.32153 &\vspace{0pt} 0.21707
&\vspace{0pt} 0.35964 &\vspace{0pt} 0.32280\\

&\vspace{0pt} 2 
&\vspace{0pt} 0.18381 &\vspace{0pt} 0.10517
&\vspace{0pt} 0.22182 &\vspace{0pt} 0.18303\\

&\vspace{0pt} 3 
&\vspace{0pt} 0.22020 &\vspace{0pt} 0.14572
&\vspace{0pt} 0.25810 &\vspace{0pt} 0.26031 \\
\hline
\end{tabular}
% }
\label{table:bass-guitar-comp}
\end{center}
\end{table}

The training procedures were performed for each type of nonlinear effect and for both instruments. Then, the models were tested with samples from the test dataset and the audio results are available online\footnote{https://github.com/mchijmma/modeling-nonlinear}. Since the \textit{mae} depends on the amplitude of the output and target waveforms, Tables \ref{table:bass-losses}-\ref{table:bass-guitar-comp} show the energy-normalized \textit{mae} for the different models when tested with various test subsets.

Table \ref{table:bass-losses} shows that the models performed well on each nonlinear audio effect task for bass guitar and electric guitar models respectively. Overall, for both instruments, \textit{model-1} achieved better results with the test datasets. For selected distortion and overdrive settings, Fig. \ref{fig:frames} shows selected input, target and output frames as well as their FFT magnitudes and spectograms. It can be seen that, both in time and frequency, the models accomplished the nonlinear target with high and almost identical accuracy. Fig. \ref{fig:ratios} shows the amplitude ratio between a test input frame and its respective target and output. It can be seen that the models were able to match precisely the input-target waveshaping curve or ratio for selected settings. The models correctly accomplished the timing settings from the nonlinear effects, such as attack and release, which are evident in the hysteresis behavior of Figs. \ref{fig:ratios}-a-b-c.

We obtained the best results with the overdrive task \#2 for both instruments. This is due to the waveshaping curves from Fig. \ref{fig:ratios}-d, where it can be seen that the transformation does not involve timing nor filtering settings. We obtained the largest error for FxChain setting \#3. Due to the extreme filtering configuration after the overdrive, it could be more difficult for the network to model both the nonlinearity and the filters. 

It is worth mentioning that the EQ task is also nonlinear, since the effects that were applied include amplifier emulation, which involves nonlinear modeling. Therefore, for this task, the models are also achieving linear and nonlinear modeling. Also, the audio samples for all the effects from the \textit{IDMT-SMT-Audio-Effects} dataset have a fade-out applied in the last 0.5 seconds of the recordings. Thus, when modeling nonlinear effects related to dynamics, this represents an additional challenge to the network. We found that the network might capture this amplitude modulation, although additional tests are required.

For the FxChain task, we evaluate the generalization capabilities of \textit{model-1} and \textit{model-2}. We test the models with recordings from different instruments (e.g. Bass guitar models tested with electric guitar test samples and vice versa). As expected, bass guitar models performed better with lower guitar notes and conversely. Also, to evaluate the performance of the models with a broader data set, we use the test subset of the NSynth Dataset \cite{engel2017neural}. This dataset consists of individual notes of 4 seconds from more than $1000$ instruments. This was done for each FxChain setting and the energy-normalized $mae$ values are shown in Table \ref{table:bass-guitar-comp}. 

It is evident that \textit{model-2} outperforms \textit{model-1} when tested with different instrument recordings. This is due to the dropout layers of \textit{model-2}, which regularized the modeling and increased its generalization capabilities. Since \textit{model-1} performed better when tested with the corresponding instrument recording, we could point towards a trade-off between optimization for a specific instrument and generalization among similar instruments. This also means the CNN and DNN layers within the models are being tuned to find certain feature patterns of the respective instrument recordings.
In other words, even though \textit{model-2} is more flexible than \textit{model-1}, the latter one is more reliable when optimizing a particular instrument. 

Other black-box modelling methods suitable for this FxChain task, such as Wiener and Hammerstein (WH) models, would require additional optimization in order to find the optimal combination of linear/nonlinear components \cite{gilabert2005wiener}. Moreover, further assumptions on the WH static nonlinearity functions (i.e. invertibility) are needed and common nonlinearities which are not invertible are for example a dead-zone and a saturation \cite{hagenblad1999aspects}. Therefore, the proposed end-to-end deep learning architecture represents an improvement of the state-of-the art in terms of flexibility, regardless of the trade-off between the two models. It makes less assumptions about the modeled audio system and is thus more suitable for generic black-box modeling of nonlinear and linear audio effects.

% The reduced number of assumptions on the audio system required in the proposed method increases its applications on generic black-box modeling of both nonlinear and linear audio effects.
% Based on the FxChain task and in comparison with black-box modeling methods such as Wiener and Hammerstein (WH), the different settings of this nonlinear task would have altered the performance of a WH model. This is because it is necessary to find the WH model with the suitable combination of linear and nonlinear components in order to obtain the best results. In addition, assumptions from the WH static nonlinearity functions are required, such as invertibility, which is a drawback when modeling much more complex nonlinearities \cite{gilabert2005wiener}. Therefore the use of the proposed end-to-end deep learning architecture makes less assumptions about the modeled audio system and is thus more suitable for generic black-box modeling of nonlinear and linear audio effects.

% \vspace{-2ex}
\section{Conclusion}
\label{sec:conclusion}

% Generalization capabilities among instruments and optimization towards an specific instrument were found among the trained models. Models with dropout layers tended to perform better with different instruments, whereas models without this type of regularization were better adjusted to the respective instrument of the training data.

% As future work, further generalization could be explored with the use of $L_1$ or $L_2$ weight regularizers as well as training data with a wider range of instruments. Also, the exploration of recurrent neural networks can improve the capabilities of the network to model much more complex audio effects. In this case, transformations involving temporal dependencies such as compression or different modulation effects. Although the model is currently running on GPU, real-time implementations could be explored.

% Possible applications for this architecture are automatic linear and nonlinear processing within an automatic mixing task. Specific instrument recordings and their respective mixing processing could be analyzed and implemented by the model.
In this work, we introduced a general purpose deep learning architecture for audio processing in the context of nonlinear modeling. Complex nonlinearities with attack, release and filtering settings were correctly modeled by the network. Since the model was trained on a frame-by-frame basis, we can conclude that most transformations that occur within the frame-size will be captured by the network. To achieve this, we explored an end-to-end network based on convolutional front-end and back-end layers, latent-space DNNs and smooth adaptive activation functions. We showed the model matching distortion, overdrive, amplifier emulation and combination of linear and nonlinear audio effects.

Generalization capabilities among instruments and optimization towards an specific instrument were found among the trained models. Models with dropout layers tended to perform better with different instruments, whereas models without this type of regularization were better adjusted to the respective instrument of the training data. As future work, further generalization could be explored with the use of weight regularizers as well as training data with a wider range of instruments. Also, the exploration of recurrent neural networks to model transformations involving long term memory such as dynamic range compression or different modulation effects. Although the model is currently running on a GPU, real-time implementations could be explored, as well as shorter input frames for low-latency applications. 

\thanks{The Titan Xp used for this research was donated by the NVIDIA Corporation.}

% \vfill\pagebreak\pagebreak

% References should be produced using the bibtex program from suitable
% BiBTeX files (here: strings, refs, manuals). The IEEEbib.bst bibliography
% style file from IEEE produces unsorted bibliography list.
% -------------------------------------------------------------------------
\bibliographystyle{IEEEbib}
\bibliography{references}

\end{document}